\documentclass[aps,prl,twocolumn,nofootinbib]{revtex4-2}

\usepackage{amsmath}
\usepackage{amssymb}
\usepackage{graphicx}
\usepackage{color}
\usepackage{gensymb}
\usepackage{float}
\usepackage{ulem}
\usepackage{empheq} % autoloads mathtols and amsmath
\usepackage{multirow,bigdelim}
\usepackage{makecell}

\newcommand{\be}{\begin{equation}}
\newcommand{\ee}{\end{equation}}
\newcommand{\bl}{\begin{align}}
\newcommand{\el}{\end{align}}
\newcommand{\bseq}{\begin{subequations}}
\newcommand{\eseq}{\end{subequations}}

\newcommand{\g}{\gamma}
\newcommand{\p}{\partial}

\renewcommand{\l}{\lambda}
\newcommand{\di}{\mathrm d}

\begin{document}
\title{Asymptotic freedom in (3+1)-dimensional projectable Ho\v rava gravity:\\ connecting ultraviolet to infrared}
\author{Andrei O. Barvinsky$^{a,b}$~, Alexander V. Kurov$^a$~, Sergey
  M. Sibiryakov$^{c,d}$\\[2mm]
{\small\it $^a$Theory Department, Lebedev Physics Institute,
Leninsky Prospect 53, Moscow 117924, Russia}\\
{\small\it $^b$Institute for Theoretical and Mathematical Physics,
Moscow State University,\\ Leninskie Gory, GSP-1, Moscow, 119991, Russia}\\
{\small\it $^c$Department of Physics \& Astronomy, McMaster
University, Hamilton, Ontario, L8S 4M1, Canada}\\
{\small\it $^d$Perimeter Institute for Theoretical Physics, Waterloo,
 Ontario, N2L 2Y5, Canada}\\
}

\begin{abstract}
We investigate the renormalization group flow of projectable Ho\v rava gravity in $(3+1)$ dimensions generated by marginal operators with respect to the Lifshitz scaling. The flow possesses a number of asymptotically free fixed points. 
We find a family of trajectories connecting one of these fixed points in the ultraviolet to the region of the parameter space where the kinetic term of the theory acquires the general relativistic form. The gravitational coupling exhibits non-monotonic behavior along the flow, vanishing both in the ultraviolet and the infrared. 
\end{abstract}

\maketitle

{\bf    Introduction} -- Ho\v rava gravity (HG)~\cite{Horava:2009uw} is a proposal to formulate quantum gravity with consistent ultraviolet (UV) limit within the realm of unitary and power-counting renormalizable quantum field theory. 
It considers a class of metric theories classically invariant at high energies under anisotropic (Lifshitz) scaling of spacetime,
\be
\label{scaling}
t\mapsto b^{-d} t\;, \qquad x^i\mapsto b^{-1} x^i\;, \quad i=1,\dots,d \:,
\ee
where $d$ is the number of spatial dimensions and $b$ is an arbitrary scaling parameter. The action of such theory contains terms with higher spatial derivatives which
improve convergence of the loop integrals, whereas higher order time derivatives are avoided ensuring that the theory remains  free from Ostrogradsky instabilities, a well-known plague of higher-derivative gravity \cite{Stelle:1976gc,Stelle:1977ry}.

For $d>1$ the scaling \eqref{scaling} does not treat time and space coordinates equally, thus breaking explicitly the Lorentz and diffeomorphism invariances. Lorentz symmetry emerges only as an approximate concept at low energies, whereas the diffeomorphism invariance is restricted to foliation preserving transformations (FDiff):
\be
\label{FDiff}
t\mapsto t'(t)\;, \qquad x^i\mapsto x'^i(t,{\bf x})\;,
\ee
where $t'(t)$ is a monotonic function.
The {\it non-projectable} version of HG \cite{Blas:2009qj}
can be sufficiently close to general relativity (GR) at low energies to pass the observational tests \cite{Blas:2014aca,EmirGumrukcuoglu:2017cfa}, albeit at the expense of some parameter tuning. 
Recent studies 
\cite{Bellorin:2022np,Bellorin:2023br}
suggest its perturbative renormalizability, but a complete proof is still pending.

In this Letter we consider a simpler {\it projectable} version which was proved to be perturbatively renormalizable in any number of spacetime dimensions~\cite{Barvinsky:2015kil,Barvinsky:2017zlx}. Its full sets of one-loop $\beta$-functions for marginal operators with respect to the scaling (\ref{scaling}) were computed in $d=2$ and $d=3$ and were shown to possess asymptotically free UV fixed points with vanishing Newton coupling \cite{Barvinsky:2017kob,towards,3+1}. 

Here we study global properties of the renormalization group (RG) flow in the space of marginal operators in the case $d=3$. In the center of our discussion is the coupling $\lambda$ appearing in the kinetic term of HG (see eq.~(\ref{Sgen})). Together with the Newton's coupling, it affects both the high- and low-energy properties of the theory. In GR $\lambda$ is forced to be unity by the full diffeomorphism invariance. By contrast, the fixed points of $d=3$ projectable HG are characterized by $\l\to\infty$. While this limit does not present any problem and corresponds to a weakly coupled theory \cite{Gumrukcuoglu:2011xg,Radkovski:2023cew}, it raises a question: Does the RG flow starting from any of the UV fixed points reach the values $\lambda\approx 1$ in the inrared (IR)? This question is non-trivial due to the presence of other couplings that may diverge before the order-one values of $\lambda$ are achieved. We answer it affirmatively by presenting a set of RG trajectories emanating from an asymptotically free fixed point at $\lambda=\infty$ and flowing  
towards $\lambda\to 1^+$ in the IR domain, with all couplings remaining regular along the flow. Interestingly, before reaching the IR region, the trajectories bypass another fixed point which is not asymptotically free. This leads to a peculiar non-monotonic behavior of the gravitational coupling along the flow.

We note that while $\lambda\to 1^+$ appears necessary for recovery of the GR limit, it is far from being sufficient. Compared to GR, projectable HG features an additional scalar excitation with a tachyonic instability at low energies \cite{Koyama:2009hc,Blas:2010hb} which presents a serious challenge to its phenomenological viability. An eventual decision on this issue will hinge on identifying the fate of this instability, or may require considerations in the strong coupling regime \cite{Mukohyama:2010xz,Izumi:2011eh,Gumrukcuoglu:2011ef}. Our results make a first bridge between the good UV properties of the theory and its IR behavior.

{\bf Projectable Ho\v rava gravity} -- We start with a brief review of HG. Preferred spacetime foliation suggests  Arnowitt--Deser--Misner (ADM) decomposition of the metric
\be
\di s^2 = N^2 \di t^2 - \gamma_{ij} (\di x^i + N^i \di t) (\di x^j +
N^j \di t)\,,
\ee
where $N$ is the lapse function, $N^i$ is the shift vector and $\g_{ij}$ is the spatial metric. These fields are assigned the following scaling dimensions with respect to \eqref{scaling}:
\be
\label{dims}
[N]=[\gamma_{ij}]=0~,~~~~~[N^i]=d-1\;.
\ee
The action invariant under FDiff and  containing only relevant and marginal operators
with respect to the anisotropic scaling reads,
\be
\label{Sgen}
S=\frac{1}{2G}\int \di t\,\di^d x\,N \sqrt{\g}\big(K_{ij}K^{ij}-\l
K^2-{\cal V}\big)\;,
\ee
where $G$ and $\l$ are marginal coupling constants, $K_{ij}=\tfrac{1}{2}\left(\p_t\gamma_{ij}-\nabla_i N_j -\nabla_j N_i\right)$ is the extrinsic curvature of the foliation, $K\equiv K_{ij}\gamma^{ij}$
and $\nabla_i$ is the covariant derivative with respect to the spatial
metric $\g_{ij}$. The potential part ${\cal V}$ depends on the $d$-dimensional metric $\g_{ij}$, its spatial derivatives and the acceleration vector $a_i = \p_i N/N$.  It does not include any time derivatives.

We focus on the {\it projectable} version of HG with spatially homogeneous lapse function which we set to be $N=1$ with the help of time reparameterizations \eqref{FDiff}. After using Bianchi identities, integration by parts and Ricci decomposition the most general expression for the potential term in $d=3$ reads \cite{Sotiriou:2009gy},
\begin{align}\label{pot31}
&{\cal V}=2\Lambda-\eta R+\mu_1 R^2+\mu_2 R_{ij}R^{ij}+\nu_1 R^3+\nu_2 RR_{ij}R^{ij}\nonumber\\
&+\nu_3R^i_jR^j_kR^k_i +\nu_4 \nabla_i R\nabla^i R+\nu_5 \nabla_iR_{jk}\nabla^i R^{jk}\;,
\end{align}
where $R_{ij}$, $R$ are the Ricci tensor and scalar of the metric $\gamma_{ij}$, and $\Lambda$, $\eta$, $\mu_a$, $\nu_b$ are the couplings.

The spectrum of perturbations propagated by this action contains a transverse-traceless (tt) graviton and a scalar mode.
For unitarity and stability in UV $G$ must be positive and $\l$ must lie in the domain:
\be
\label{lambdaunitary}
\lambda<1/3~~\text{or}~~\lambda>1\;.
\ee

If $\Lambda=0$, the theory possesses a flat background solution with the dispersion relations of the perturbations around it,
\bseq
\label{disp}
\begin{align}
\label{disp1}
&\omega_{tt}^2=\eta k^2+\mu_2 k^4+\nu_5k^6\;,\\
\label{disp2}
&\omega_s^2=\frac{1-\l}{1-3\l} \big(-\eta k^2+(8\mu_1+3\mu_2)k^4\big)+u_s^2\nu_5k^6\;,
\end{align}
\eseq
where $u_s=\sqrt{\frac{1-\l}{1-3\l}\left(\frac{8\nu_4}{\nu_5}+3\right)}$.
For consistency  of the UV limit, $\nu_5$ and $u_s$ must be positive. The negative sign in front of the $\eta k^2$ term in (\ref{disp2}) signals an instability of the flat background with respect to the long scalar modes (we assume $\eta>0$ for the stability of the (tt) gravitons). 
Elucidating the fate of this instability is beyond the scope of this Letter.

The instability gets removed if we restrict to the high energy limit by keeping only the last five marginal terms in the potential \eqref{pot31}. 
The essential couplings, i.e. the couplings whose $\beta$-functions do not depend on the gauge, can be chosen 
as \cite{towards}
\be \label{new_couplings}
{\cal G} = \frac{G}{\sqrt{\nu_5}},\quad \lambda,
\quad u_s, \quad v_a = \frac{\nu_a}{\nu_5}, \quad   a=1,2,3.
\ee
The one-loop beta functions for all essential marginal couplings \eqref{new_couplings} were derived in \cite{towards, 3+1} and have the form,
%\begin{widetext}
\begin{subequations}\label{betafun}
 \begin{align}
    \beta_\lambda  &=  {\cal G}\frac{27(1\!-\!\l)^2+3u_s(11\!-\!3\l)(1\!-\!\l)-2u_s^2(1\!-\!3\l)^2}{120\pi^2 u_s (1+u_s)(1\!-\!\l)}\;, 
    %+ O({\cal G}^2), 
    \label{beta_lam} \\
    \beta_{\cal G}  &= {\cal G}^2\frac{ 1}{26880\pi^2(1-\lambda)^2(1-3\lambda)^2
    (1+u_s)^3 u_s^3} \notag\\
    &\qquad~\times\sum_{n=0}^7 u_s^n\, {\cal P}^{\cal G}_n[\l,v_1,v_2,v_3] \;,
    %+ O({\cal G}^3),  
    \label{betaG} \\
    \beta_\chi    &={\cal G} \frac{ A_\chi  }{26880\pi^2(1-\lambda)^3(1-3\lambda)^3
    (1+u_s)^3 u_s^5} \notag\\
    &\qquad~\times\sum_{n=0}^9 u_s^n\, {\cal
    P}^{\chi}_n[\l,v_1,v_2,v_3] \;,
    %+ O({\cal G}^2), 
    \label{beta_chi}
\end{align}
\end{subequations}
%\end{widetext}
where we collectively denoted $\chi= (u_s,v_1,v_2,v_3)$ and the coefficients $A_\chi$ are equal to
$A_{u_s} = u_s (1-\lambda)$, $A_{v_1} = 1$, $A_{v_2} =A_{v_3} = 2$.
Note that the coupling ${\cal G}$ controlling the overall strength of gravitational interactions factorizes.
The functions ${\cal P}^{\cal G}_n$, ${\cal P}^{\chi}_n$ are polynomials in $\l$ and $v_a$ with integer coefficients. ${\cal P}^{\cal G}_n$, ${\cal P}^{u_s}_n$ and ${\cal  P}^{v_a}_n$ are respectively
of the fourth, fifth and sixth order in $\lambda$. The maximum overall power of the couplings $v_a$ is two for ${\cal P}^{\cal G}_n$, ${\cal P}^{u_s}_n$ and three for  ${\cal  P}^{v_a}_n$. Explicit expressions for these polynomials are cumbersome and can be found in \cite{3+1}.

{\bf Fixed points} -- Ref.~\cite{3+1} has found several fixed points at finite values of $\lambda<1/3$ in the left part of the domain 
\eqref{lambdaunitary}. Since the RG flow cannot cross non-unitary region, the trajectories starting from these fixed points in UV cannot reach the domain
$\l\to1^+$ in IR. We performed a numerical search for fixed points with finite $\l>1$ and obtained null results (the detail will be reported in \cite{4dRGlong}). In what follows we focus on fixed points at $\l\to\infty$. This limit was proved to be regular  and independent on the direction  $\l\to\pm \infty$, at least in perturbation theory  \cite{Radkovski:2023cew}.

To analyze it, we introduce a new variable, 
\be\label{rho}
\varrho \equiv \frac{3(1-\l)}{1-3\l}\;, \qquad \l = \frac{3-\varrho}{3(1-\varrho)}\;.
\ee
In the unitary domain \eqref{lambdaunitary} the coupling $\varrho$ is positive. The limit $\l=\infty$ corresponds to finite $\varrho=1$, whereas $\l=1$ corresponds to $\varrho=0$. 
$\beta$-function of $\varrho$,
\be\label{betarho}
\beta_\varrho =
3{\cal G}(1-\varrho)\frac{2u_s^2+u_s\varrho(4-5\varrho)-3\varrho^2}{40\pi^2u_s(1+u_s)\varrho},
\ee
vanishes for $\varrho=1$ and all other $\beta$-functions (\ref{betaG}), (\ref{beta_chi}) are regular, implying regularity of the RG flow in the parameterization $(\varrho, u_s, v_a)$.

All fixed points of RG flow in the hyperplane $\varrho=1$ are listed in Table \ref{tabFP2}.
First three fixed points are attractive along $\varrho$ direction, implying that any RG trajectory starting from them will stay in the $\varrho=1$ hyperplane. Such trajectories are not of interest to us.  Fixed points \textnumero 6 and 8 are not asymptotically free. This leaves only 
the fixed points 
\textnumero 4, 5 and 7 as possible origin for trajectories 
connecting the regions with infinite and finite 
$\l$. A numerical investigation has shown that the trajectories emanating from the points \textnumero 4 and 7 run into singularities with divergent couplings $u_s$ or $v_a$, before $\varrho$ undergoes any essential deviation from $1$ (details will be reported elsewhere \cite{4dRGlong}). 
Here we focus on trajectories emanating from the fixed point \textnumero 5 which we label as A below. We will see that these trajectories are also affected by the fixed point \textnumero 6 which we label as B.

\begin{table}[h]
%\begin{center}
\begin{tabular}{|c| c | c | c | c | c | c |c |}
 \hline
\textnumero&$u_s$  & $v_1$ & $v_2$ & $v_3$ &   $\beta_{\cal G}/{\cal G}^2$& AF?
 &\makecell{Can flow\\out of \\ $\varrho=1$?}\\ [0.5ex]
\hline\hline
1&0.0195& 0.4994 & -2.498 & 2.999 &  -0.2004 &yes&no\\ [0.5ex]
\hline
 2&0.0418 & -0.01237 & -0.4204 & 1.321 &  -1.144&yes&no \\ [0.5ex]
\hline
3&0.0553 & -0.2266 & 0.4136 & 0.7177 &  -1.079 &yes&no\\ [0.5ex]
 \hline
4&12.28 & -215.1 & -6.007 & -2.210 &   -0.1267 &yes&yes\\ [0.5ex]
\hline
5&21.60 & -17.22 & -11.43 & 1.855 &   -0.1936 &yes&yes\\ [0.5ex]
 \hline
6&440.4 & -13566 & -2.467 & 2.967 &   0.05822 &no&yes\\ [0.5ex]
 \hline
7&571.9 & -9.401 & 13.50 & -18.25 &  -0.0745 &yes&yes\\ [0.5ex]
 \hline
8&950.6 & -61.35 & 11.86 & 3.064 &   0.4237 &no&yes\\ [0.5ex]
 \hline
  \end{tabular}
    \caption{Fixed points of projectable HG at $\varrho=1$. Listed data are: the values of the couplings $\chi$ at a fixed point; the value of the $\beta$-function for the  coupling $\cal G$; whether the point is asymptotically free or not; whether the trajectories starting from the point can flow out of the hyperplane $\varrho=1$.}
    \label{tabFP2} 
%\end{center}
\end{table}

{\bf RG flow } --  $\beta$-functions \eqref{betafun} are defined as derivatives of the couplings $g_i$ with respect to   
$\log k_*$, the logarithm of the sliding momentum scale. 
%It is related to the energy scale as
%$\log k_* = \frac13 \log E_*$ where the %$\frac13$-factor is due to the Lifshitz %scaling. 
It is convenient to change the parameterization of the RG trajectories by introducing new independent variable 
$\tau$ through $d\tau = {\cal G}\, d\log k_*$. Defining $\tilde \beta_{g_i}=dg_i/d\tau$, we find that the RG flow in the subspace $(\varrho,u_s,v_a)$ decouples from the running of ${\cal G}$. 

In the vicinity of a fixed point, the RG flow can be analyzed with the help of the stability matrix $B_i^{\:\:j}$,
\be\label{stabmat}
\tilde{\beta}_{g_i} \cong \sum_j B_i^{\:\:j} ( g_j- g_j^*), \quad B_i^{\:\:j} \equiv   \left(\frac{\partial \tilde{\beta}_{g_i}}{\partial g_j}\right)\Big|_{g=g^*},
\ee
where $g_i^*$ are fixed point values of the couplings. The eigenvalues $\theta^J$ of the stability matrix $B_i^{\:\:j}$ determine whether the RG flow is attracted to (${\rm Re}\, \theta^J <0$) or repelled from (${\rm Re}\, \theta^J >0$) the fixed point along the eigendirection as the energy scale increases. When the energy is lowered, the situation is opposite: The flow is attracted to the fixed point for ${\rm Re}\, \theta^J >0$ and is repelled from it for ${\rm Re}\, \theta^J <0$. Below we use the terms attraction/repulsion in the latter sense of running from UV to IR.

Since $\tilde{\beta}_\varrho$ is proportional to $(1-\varrho)$, all its derivatives, except the one with respect to $\varrho$, vanish 
at $\varrho=1$. Hence the matrix $B_i^{~j}$ has a single non-zero element in the row corresponding to ${\varrho}$. This element is situated on the diagonal giving an eigenvalue $\theta^1=B_{\varrho}^{~\varrho}= d\tilde{\beta}_\varrho/d \varrho\,|_{g=g^*}$. 
The corresponding eigenvector has non-zero $\varrho$-component, while the $\varrho$-components of all other eigenvectors vanish. The sign of $\theta^1$
indicates whether the fixed point is attractive ($\theta^1>0$) or repulsive ($\theta^1<0$) along the $\varrho$-direction.  The eigenvalues of the stability matrix in the variables $(\varrho, u_s, v_a)$ for points A and B are provided in Table \ref{EVlam}.
\begin{table}[h]
\begin{center}
\begin{tabular}{| c  || c | c | c | c |c|}
 \hline
Fixed point& $\theta^{1}$ & $\theta^{2}$ & $\theta^{3}$ & $\theta^{4}$& $\theta^{5}$  \\ [0.5ex]
\hline\hline
A&-0.0141&-0.0700 & 0.257&0.320 & 0.0657 \\ [0.5ex]
 \hline
B&-0.0151&0.603& 0.308& \multicolumn{2}{c|}{0.092 $\pm$ 0.289 $i$} \\ [0.5ex]
 \hline
  \end{tabular}
    \caption{Eigenvalues $\theta^I$ of the stability matrix for the fixed points A and B.}
    \label{EVlam}
\end{center}
\end{table}

%Note that the point B has a pair of complex conjugate eigenvalues. This may appear puzzling since the eigenvalues of the stability matrix are commonly associated with the anomalous dimensions of the operators driving the RG flow which are  expected to be real in a unitary theory. However, this reasoning breaks down for asymptotically free gauge theories due to the mismatch between the full non-linear gauge group along the flow and its linearized version at the fixed point. While the deformation operators are invariant under the former, they are not under the latter and thus do not need to obey unitarity constraints of the (free) conformal theory at the fixed point. This subtlety deserves further study.

To construct RG trajectories starting from a fixed point $g^*$ in the UV, we slightly shift the initial condition away from $g^*$ in the repulsive direction and numerically integrate the equations
 \be\label{RGeq}
 \left\{
\begin{aligned}
&\frac{d g_i}{d \tau} = \tilde{\beta}_{g_i},   \quad g_i = (\varrho,u_s,v_1,v_2,v_3),\\
&g_i (0) = g_i^* + \varepsilon\, c_J\, w_i^{J} 
\end{aligned}
\right.
\ee
from $\tau=0$ towards $\tau\to-\infty$. 
Here $\varepsilon$ is a small parameter, $c_J$ are constants satisfying
$\Sigma_J (c_J)^2=1$, 
and $w^J_i$ are eigenvectors enumerated by the index $J$, $B_i^{\:\:j} w^J_j = \theta^J w^J_i$, with $\theta^J<0$.
The components of two repulsive eigenvectors $A1,A2$ for the point A and  the unique repulsive eigenvector $B1$ for the point B, are given in  Table~\ref{EVlam5}. 
%The trajectory is independent of the choice of $\varepsilon$ when it is sufficiently small.

\begin{table}[h]
\begin{center}
\begin{tabular}{| c  || c | c | c | c | c |}
 \hline
\makecell{Eigen-\\vector}  & $w_\varrho$& $w_{u_s}$ & $w_{v_1}$  & $w_{v_2}$ & $w_{v_3}$   \\ [0.5ex]
\hline\hline
$A1$& 0.0423&0.998 &
-0.0398 & 5.25$\times10^{-3}$ & 5.57$\times10^{-3}$ \\ [0.5ex]
 \hline
 $A2$ & 0& -0.967 &
 -0.115 & -0.224 & 0.0480 \\ [0.5ex]
 \hline
 $B1$& 2.19$\times10^{-5}$&0.0162 &
 -0.999 & 1.87$\times10^{-5}$ & 5.69$\times10^{-6}$ \\ [0.5ex]
 \hline
 \end{tabular}
    \caption{Repulsive eigenvectors of the stability matrix for the fixed points A and B.\label{EVlam5}}
\end{center}
\end{table}

\begin{figure}[h]
\centerline{\includegraphics[width=0.48 \textwidth]{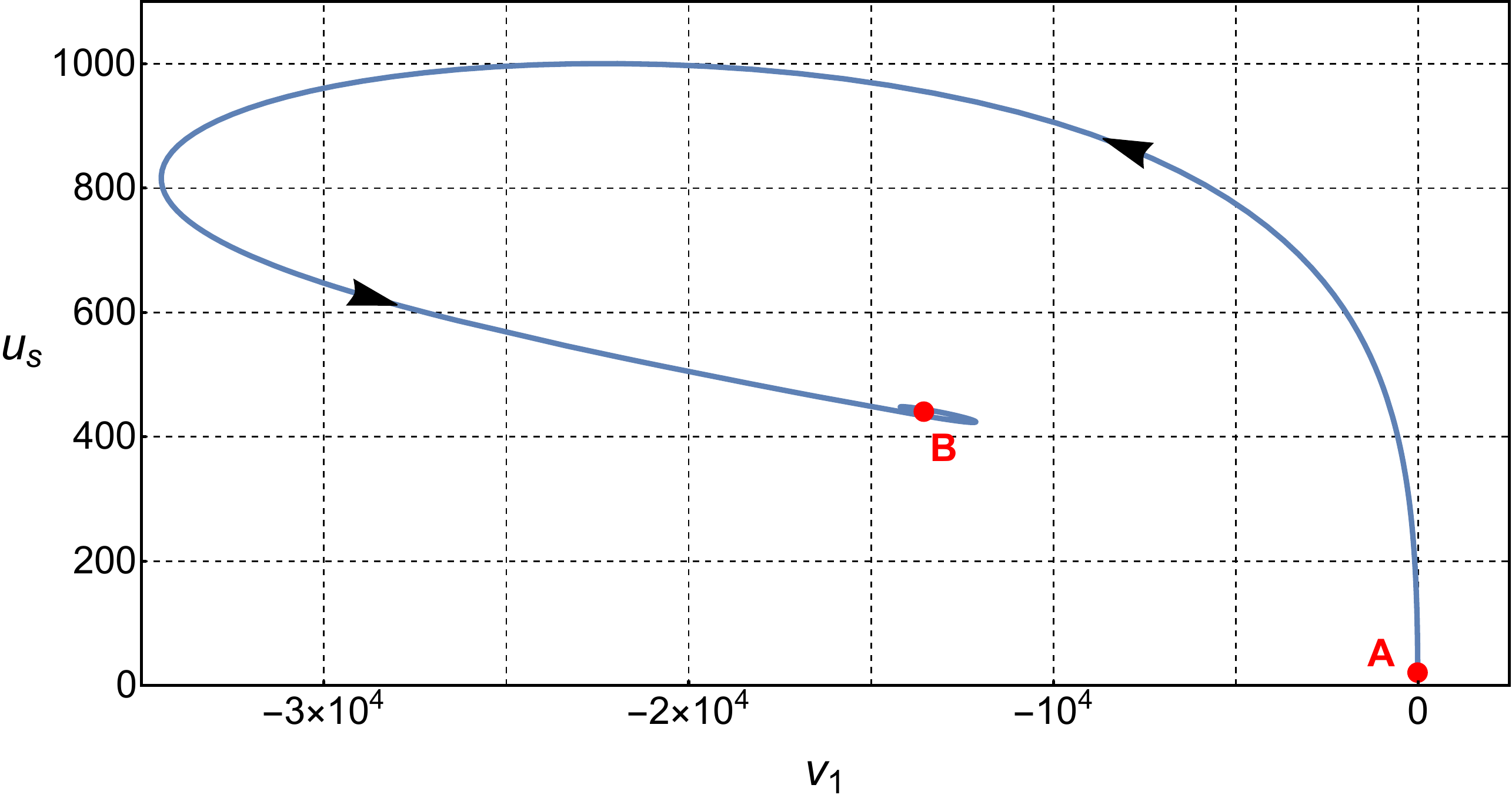}}
~\\
\includegraphics[width=0.48 \textwidth]{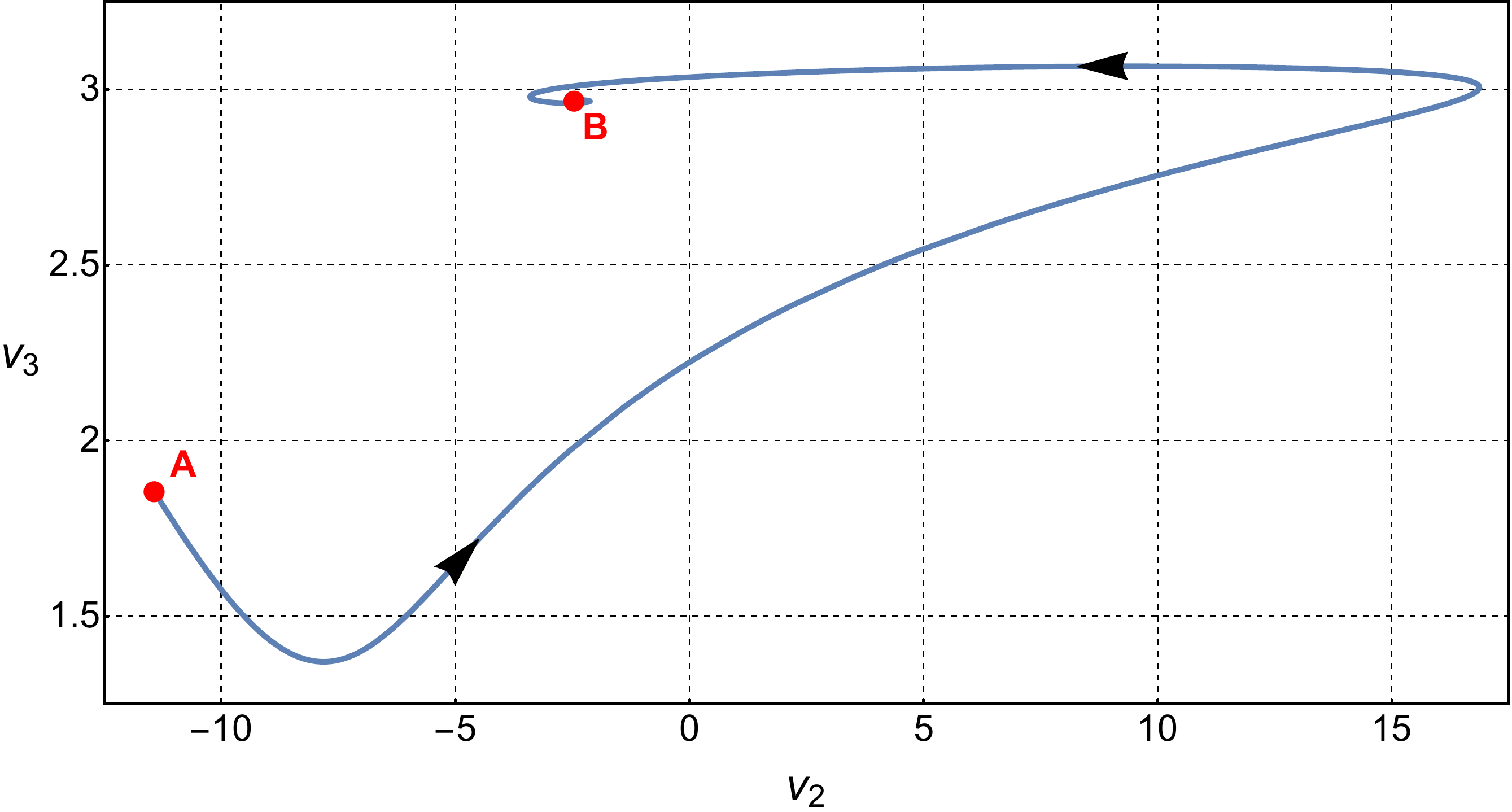}
\caption{RG trajectory connecting fixed points A and B. 
The trajectory lies entirely in the hyperplane $\varrho=1$. Panels show its projections on the $(u_s,v_1)$ and $(v_2,v_3)$ planes. 
Arrows indicate the flow from UV to IR.\label{AtoB}}
\end{figure}

{\it From A to B} -- First we build the trajectory flowing from point A along the  eigenvector $A2$. Since this vector has zero $\varrho$-part, the trajectory stays in the hyperplane $\varrho=1$. Notably, if we take $\varepsilon<0$, the trajectory arrives at the point B, see Fig.~\ref{AtoB}.  
In the opposite case, $\varepsilon>0$, the couplings $u_s,v_a$ diverge at finite $\tau$, i.e. the trajectory hits a singularity. 

The existence of a trajectory connecting the points A and B may seem surprising. However,  we notice that all eigenvalues of the stability matrix at point B, except $\theta^1$, have positive real part and thus 
B is absolutely IR attractive in the hyperplane $\varrho=1$. It happens that the point A lies on the boundary of its basin of attraction.

{\it From B to $\l\to1^+$} -- Let us now look closer at the point B. It has a unique repulsive direction, pointing away from the $\varrho=1$ hyperplane, see
Tables \ref{EVlam} and \ref{EVlam5}. This gives rise to two RG trajectories, depending on the sign of $\varepsilon$ in the initial conditions \eqref{RGeq}. 
%with the initial condition of %the form
%\be
%g_i(0) = g_i^{*(B)}  +  %\varepsilon\, w^{B1}_i 
%\ee

\begin{figure}[h]
\centering
\includegraphics[width=0.48 \textwidth]{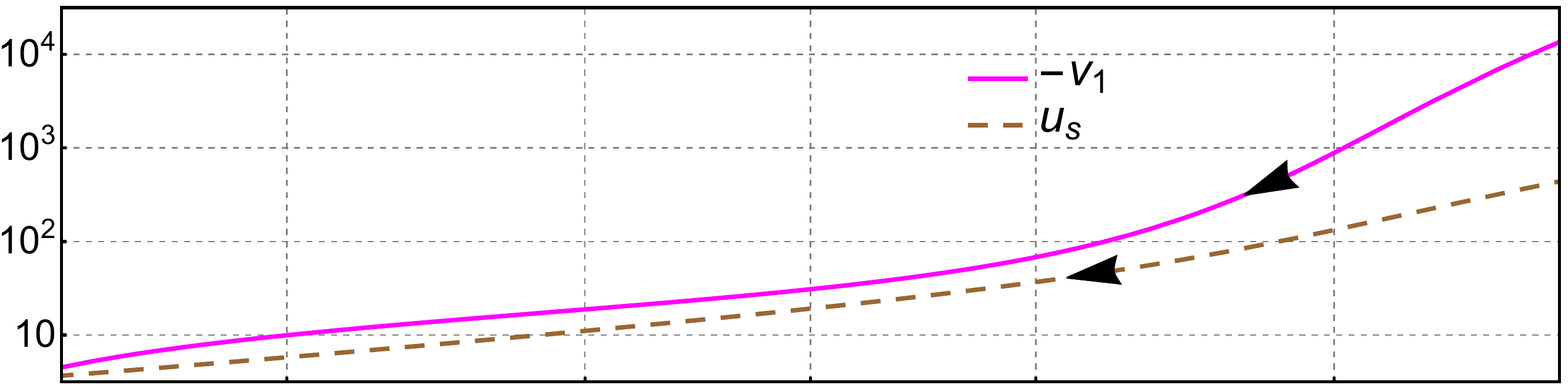}
\includegraphics[width=0.48 \textwidth]{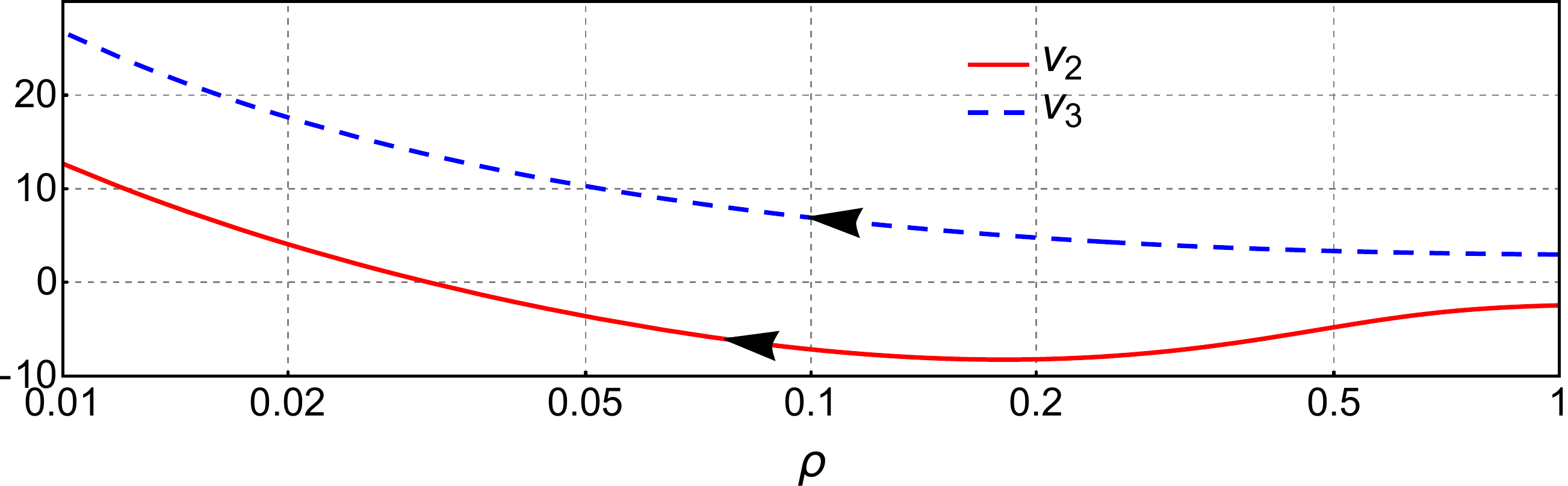}
\caption{The couplings $(u_s,v_a)$ as functions of $\varrho$ along the RG trajectory from the fixed point B to $\varrho=0$ ($\l\to1^+$). %Note the logarithmic scale in the upper panel.
Arrows indicate the flow from UV to IR.}
\label{Btolam1+}
\end{figure}

On the solution with $\varepsilon<0$, the coupling $\varrho$ monotonically decreases and at $\tau\to-\infty$ reaches the boundary of the unitary domain \eqref{lambdaunitary} $\varrho\to0$ ($\l\to1^+$).  The behaviour of other couplings is shown in Fig.~\ref{Btolam1+}. 
The couplings $v_a$ approach some finite values of order $O(1)$ or $O(10)$ before they start to rapidly grow in the small vicinity of $\varrho=0$ (not shown in the plot). This divergence can be attributed to the presence of large inverse powers of $(1-\lambda)$ in the beta functions \eqref{beta_chi}. The coupling $u_s$ tends to zero when $\varrho\to0$. 
%\ss{are we sure?}

On the trajectory with $\varepsilon>0$ the coupling $\varrho$ monotonically increases and at $\tau\to-\infty$ reaches another boundary of the unitary domain \eqref{lambdaunitary} $\varrho\to\infty$ ($\l\to1/3^-$). 

Thus, the RG trajectories starting from B are regular and span the whole unitary domain (\ref{lambdaunitary}). The point B, however, is not asymptotically free, see Table~\ref{tabFP2}, and cannot serve as a UV limit of a perturbative RG flow. Still, it allows us to construct a regular flow from UV to IR.

{\it From A to $\l\to1^+$} -- To this end, we consider a general linear combination of vectors $A1$ and $A2$ in the initial  condition (\ref{RGeq}) at the point A. If we set $c_{A2}=0$, the resulting trajectory runs into a singularity with $v_1$ diverging to negative infinity at finite value of RG parameter $\tau$. But if $c_{A2}$ is
larger than a certain critical value $\sim 2\times 10^{-3}|\varepsilon|$,  the trajectory passes in the neighborhood of the point B and gets attracted to the trajectory shown in Fig.~\ref{Btolam1+}. 
This gives the sought-after family of RG trajectories starting from the asymptotically free UV fixed point A and running into the region $\l\to 1^+$ in IR.

\begin{figure}[h]
\centering
\includegraphics[width=0.48 \textwidth]{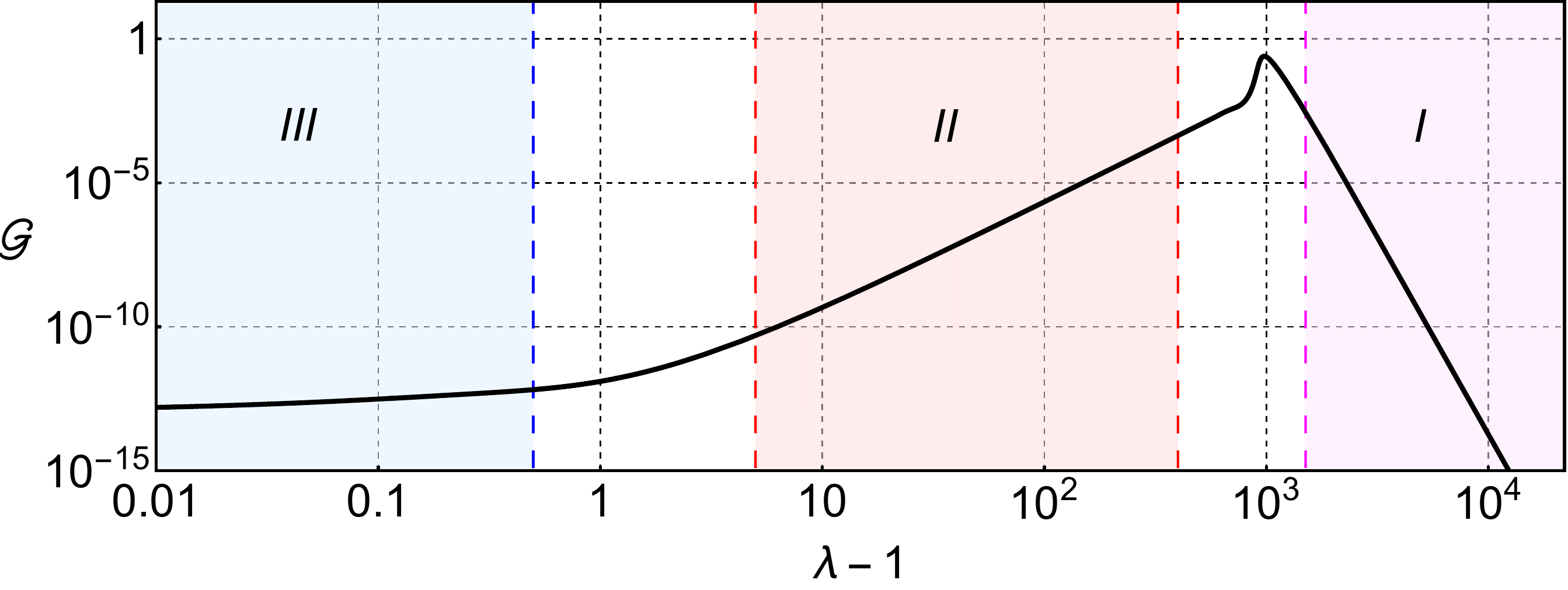}
\caption{Behaviour of ${\cal G}$ as a function of $(\l-1)$ along an RG trajectory connecting the point A to $\l\to 1^+$. In regions $I$, $II$ and $III$ the dependence is well described by the power law ${\cal G}\propto (\l-1)^k$ 
with 
$k_I=-13.69$, $k_{II}=3.84$, $k_{III}\approx 0.37$. %\ss{interchange labels $I\leftrightarrow III$ in the plot, shade the regions}
}
\label{PlotG}
\end{figure}

{\bf The behaviour of $\boldsymbol{\cal G}$} -- The RG equation for the overall coupling ${\cal G}$ has the form
$d {\cal G}/d \tau  =  {\cal G} \hat{\beta}_{\cal G}$,
where $\hat\beta_{\cal G}$ is ${\cal G}$-independent. This can be easily integrated: 
\be\label{Gsol}
{\cal G}(\tau) = {\cal G}(0) \exp{\int_{0}^\tau d \tau'\, \hat{\beta}_{\cal G}(\tau')} .
\ee
In Fig.~\ref{PlotG} we plot ${\cal G}$ as a function of $\lambda$ on an RG trajectory belonging to the flow from A to $\lambda\to 1^+$. We see a non-monotonic behavior which has a transparent explanation. When $\l$ decreases from infinity, ${\cal G}$ first grows as the trajectory approaches the strongly coupled fixed point B and then goes down when the trajectory leaves the vicinity of B. Its behavior in regions $I$, $II$ and $III$ is well described by the power law ${\cal G}\propto (\l-1)^k$. The regions $I$ and $II$ are dominated by the fixed points A and B, respectively. The corresponding exponents are obtained by considering 
\be
\frac{d {\cal G}}{d\l} = \frac{\beta_{\cal G}}{\beta_\l}\approx \frac{{\cal G}}{\l} \cdot\frac{\hat\beta_{\cal G}}{\hat\beta_\l}\bigg|_{A,B},
\ee 
where $\hat\beta_\l\equiv\beta_\l/\l$ is finite at $\l\to\infty$, and the ratio of the $\beta$-functions in the last expression is evaluated at the points A or B. This yields $k_I=(\hat\beta_{\cal G}/\hat\beta_\l)|_A=-13.69$ and $k_{II}=(\hat\beta_{\cal G}/\hat\beta_\l)|_B=3.84$. For the region $III$ we fit the exponent numerically, with the result $k_{III}\approx0.37$. 

The coupling ${\cal G}$ reaches maximum in the vicinity of the point B. This maximum must be less than one for the consistency of the perturbative expansion. Since ${\cal G}$ steeply decreases as the flow goes from B towards IR, the IR value of ${\cal G}$ happens to be very small. It depends on the coefficients of the vectors $A1$, $A2$ in the initial conditions at the point A which control how closely the trajectory bypasses the point B. Typically, we found ${{\cal G} \sim 10^{-13}}$ at $\l\approx 1.01$. It is possible to reach ${\cal G}$ as high as $\sim 10^{-4}$, but only at the expense of an extreme fine tuning of the initial conditions at point A.
This suggests a large natural hierarchy between the effective low-energy Planck mass $M_{Pl}\sim G^{-1/2}$ and the Lorentz violating scale ${M_{LV}\sim \nu_5^{-1/4}}$ in this scenario.   

The fact that ${\cal G}$ decreases towards IR may suggest that in IR the theory becomes free again. This is unlikely, since the inverse powers of $(1-\l)$ in the $\beta$-functions (\ref{betafun}) are expected to jeopardize the perturbative expansion. Moreover, a complete study of the IR limit should take into account the relevant operators which we disregarded in this work. We leave this analysis for future.

{\bf Acknowledgements} -- We thank Diego Blas, Davide Gaiotto, Ted Jacobson, Shinji Mukohyama, Niayesh Afshordi, Maxim Pospelov, Oriol Pujolas, Jury Radkovski, Marc Schiffer for useful discussions.
The work of A.B. and A.K. was supported by the Russian Science Foundation grant No 23-12-00051. The work of S.S. is supported by the Natural Sciences
and Engineering Research Council (NSERC) of Canada. Research at Perimeter Institute is supported in part by the Government of Canada through the Department of Innovation,
Science and Economic Development Canada and by the Province of Ontario through the
Ministry of Colleges and Universities.

\bibliography{FP}

\end{document}